\documentclass{emulateapj}

\slugcomment{To appear in the {\em Astrophysical Journal Letters}}
\shorttitle{Intermediate-Mass Black Hole in NGC\,4395}
\shortauthors{Wrobel \& Ho}
\begin{document}
\title{Radio Emission on Sub-Parsec Scales from the
       Intermediate-Mass Black Hole in NGC\,4395}
\author{J. M. Wrobel\altaffilmark{1} and L. C. Ho\altaffilmark{2}}

\altaffiltext{1}{National Radio Astronomy Observatory, P.O. Box O,
Socorro, NM 87801; jwrobel@nrao.edu}

\altaffiltext{2}{The Observatories of the Carnegie Institution of
Washington, 813 Santa Barbara Street, Pasadena, CA 91101;
lho@ociw.edu}

\begin{abstract}
The Seyfert 1 nucleus of NGC\,4395 is energized by a black hole of
mass $3.6 \times 10^5 M_\odot$ (Peterson et al.), making it one of
only two nuclear black holes of intermediate mass, $10^3 - 10^6
M_\odot$, detected in the radio regime.  Building upon UV and X-ray
evidence for outflows from this Seyfert nucleus, the VLBI High
Sensitivity Array was used at 1.4~GHz to search for extended structure
on scales greater than 5~mas (0.1~pc).  Elongated emission was
discovered, extending over 15~mas (0.3~pc) and suggesting an outflow
on sub-parsec scales from this intermediate-mass black hole.  The
Seyfert nucleus is located at the center of an elliptical star
cluster, and the elongation position angle of the sub-parsec radio
structure is only 19\arcdeg\, from the star cluster's minor axis.
\end{abstract}


\section{Motivation}

Dynamical studies have established that supermassive black holes, with
masses $10^6 - 10^9 M_\odot$, occur in the nuclei of most nearby
galaxies with stellar bulges.  The focus has now turned to searches
for intermediate-mass black holes (IMBHs), with masses $10^3 - 10^6
M_\odot$, in nearby galactic nuclei.  Finding IMBHs will help to
define the nature of the seed black holes that figure in models for
the growth of black holes over cosmic time and to predict the
gravitational radiation background expected for {\em LISA\/} due to
mergers of IMBHs.  In this mass regime, dynamical searches currently
fail beyond the Local Group and are being replaced by surveys for
signatures of active galactic nuclei \citep[AGN;][]{gre04}.  Such AGN
searches are strongly guided by the discovery of a candidate IMBH in
\object{NGC 4395} \citep{fil03}, recently established as a bona fide
nuclear IMBH with mass (3.6$\pm$1.1)$\times 10^5 M_\odot$
\citep{pet05}.

NGC\,4395 is a bulgeless Sdm galaxy at a Cepheid distance of
4.3$\pm$0.3~Mpc with a scale 0.021~pc~mas$^{-1}$ \citep{thi04}.  Its
nucleus has the emission properties of a Seyfert 1, with broad
permitted optical and UV emission lines \citep{fil93,fil03,pet05} and
a pointlike, hard X-ray source that is highly time-variable
\citep{shi03,mor05,vau05}.  The reverberation study by \citet{pet05}
reports a characteristic diameter for the UV broad-line-region (BLR)
of 7.0$\times 10^{-5}$~pc or 0.0033~mas.  There are hints of an
outflowing UV absorber close to the nucleus of this AGN \citep{cre04}.
Also, in luminous AGNs, warm X-ray absorbers close to the nucleus are
often associated with outflows; in NGC\,4395 such absorbers have been
inferred from variations in their ionization \citep{shi03} and/or
column densities \citep{mor05}.  While these X-ray and UV hints of
outflows are intriguing, seeking structural evidence for an outflow is
an essential next step.

NGC\,4395 is one of only two nuclear IMBHs that have been detected in
the radio regime \citep{gre06}, important because VLBI on mas scales
can then be used to search for structural evidence for an outflow on
sub-parsec scales.  At 1.4~GHz, the Very Long Baseline Array
\citep[VLBA;][]{nap94} detected one source, with flux density
0.53$\pm$0.13~mJy, spectral power $1.2 \times 10^{18}$~W~Hz$^{-1}$,
diameter less than 11~mas (0.2~pc), and brightness temperature more
than 2 million K \citep{wro01}.  While these traits are consistent
with the VLBA detection being accretion-powered by the IMBH, the
resolution and sensitivity of the VLBA image failed to provide
quantitative structural information.  This Letter reports improved
imaging of the Seyfert nucleus of NGC\,4395 at 1.4~GHz, using a VLBI
High Sensitivity Array (HSA) consisting of the VLBA, the Very Large
Array \citep[VLA;][]{tho80} operating in its phased mode, and the
Robert C. Byrd Green Bank Telescope
\citep[GBT;][]{jew04}.\footnote{The VLBA, the VLA, and the GBT are
operated by the National Radio Astronomy Observatory, which is a
facility of the National Science Foundation, operated under
cooperative agreement by Associated Universities, Inc.}  The HSA
imaging is described in Section 2.  The implications for the Seyfert
nucleus are explored in Section 3, with an emphasis on the discovery
of elongated structure on sub-parsec scales.

\section{HSA Imaging}

The HSA was used to observe NGC\,4395 and calibrators on 2005 May 1
0000-0800 UT.  Data were acquired in dual circular polarizations with
4-level sampling and at a center frequency of 1.43849~GHz with a
bandwidth per polarization of 32~MHz, formed from four contiguous
baseband channels each of width 8~MHz.  Antenna separations spanned
50-8600 km.  Phase-referenced observations were made in the nodding
style.  A 200 s observation of NGC\,4395 was preceded and followed by
a 100 s observation of the phase, rate, and delay calibrator
J1220+3431 about 1.5\arcdeg\, from NGC\,4395.  Observation and
correlation assumed a coordinate equinox of 2000.  All {\em a
priori\/} calibrator positions were taken from the Goddard VLBI global
solution 2004 f,
\footnote{http://gemini.gsfc.nasa.gov/solutions/2004f\_astro/} and had
one-dimensional errors at 1 $\sigma$ better than 1 mas.  The {\em a
priori\/} position for NGC\,4395 was from \citet{wro01} after
adjusting for the solution 2004 f position for J1220+3431.

Data editing, calibration, and imaging were done using the 2005
December 31 release of the NRAO AIPS software, following strategies
outlined in the AIPS
Cookbook\footnote{http://www.aoc.nrao.edu/aips/cook.html}.
Correlation was affected by inappropriate Earth Orientation
Parameters, so corrections to those parameters were made
\citep{wal05}.  Data deletion was based on system flags recorded at
observation and media weights recorded at correlation.  Data
compromised by radio frequency interference were also deleted, as were
data on baselines involving any antenna observing below an elevation
of 20\arcdeg.  For NGC\,4395, these steps yielded a total integration
time of 4 hr.  Corrections for the dispersive delays caused by the
Earth's ionosphere were made using electron-content models from the
Jet Propulsion Laboratory.  Ancillary HSA data were used to set the
amplitude scale to an accuracy of about 5\%, after first correcting
for sampler errors.  The visibility data for J1220+3431 were used to
generate phase-referenced visibility data for NGC\,4395 and for the
check source J1215+3448, observed to assess both the differential
astrometry and the coherence losses.  All calibrators were
phase-self-calibrated.  No self-calibrations were performed on
NGC\,4395.  No polarization calibration was performed, as only an
upper limit on the linear polarization percentage of NGC\,4395 was
sought.  AIPS task {\em imagr\/} was used to image all calibrators in
Stokes $I\/$ and NGC\,4395 in Stokes $I\/$, $Q\/$, and $U\/$.

The visibility data for NGC\,4395 were imaged with natural weighting,
achieving the expected theoretical rms noise within the field-of-view
limits set by time and bandwidth averaging.  The most constraining
limit follows from accepting, at the field edge, a 5\% drop in the
peak amplitude due to averaging over each 8 MHz baseband channel; the
resulting field-of-view diameter is about 590~mas.  A region of
diameter 550~mas (12~pc) was searched in Stokes $I\/$ to match the
upper limit of the unresolved VLA detection \citep{ho01}.  The Seyfert
nucleus was easily detected above 5 $\sigma$ in the cleaned image and
found to be slightly extended, with an integrated flux density of
0.74$\pm$0.04~mJy being recovered (Fig.~1, left).  There were also
hints of 5 $\sigma$ emission about 250~mas (5.2~pc) to the west of the
nucleus.  Nothing else was detected in Stokes $I\/$ above 5 $\sigma$
or a brightness temperature threshold of 250,000~K.  Nothing was
detected in Stokes $Q\/$ or $U\/$ so those images were not cleaned.
Near the Seyfert nucleus, the image of linearly polarized intensity $P
= \sqrt{Q^2 + U^2}$ revealed no emission above 5 $\sigma$, implying an
upper limit of 2\% in the polarization percentage.

The visibility data for NGC\,4395 were also imaged with a
``robustness'' parameter of -0.3 for an optimal balance between
resolution and sensitivity \citep{bri95}.
Guided by the 4 $\sigma$ contours in the
cleaned Stokes $I\/$ image (Fig.~1, right), emission extends from the
peak for about 15~mas (0.3~pc) along a position angle (PA) of about
28\arcdeg.  Analysis of the check source J1215+3448, observed
0.9\arcdeg\, from J1220+3431 and imaged with the same robustness,
implied a correction of about 2\% for residual errors in the phase
calibration and a one-dimensional error of about 3 mas in the
differential astrometry.  NGC\,4395 was observed 1.5\arcdeg\, from
J1220+3431, so for that still-desirable geometry these corrections are
conservatively doubled to about 4\% and 6 mas, respectively.  A
quadratic fit to the peak in the right panel of Figure~1 gave a
corrected value of 0.24~mJy~beam$^{-1}$ and a position of
$\alpha(J2000) = 12^{h} 25^{m} 48^{s}.8774$ and $\delta(J2000) =
33^{\circ} 32' 48''.715$, with a one-dimensional error of 6~mas (the
quadratic sum of the errors due to the signal-to-noise ratio, in the
position for J1220+3431, and in the differential astrometry).

A standard analysis of the data from the internal VLA baselines
yielded an integrated flux density of 1.69$\pm$0.08~mJy with a
deconvolved diameter of less than 2\arcsec\, (42~pc), plus no evidence
for adjacent emission on larger scales.  For comparison, \citet{ho01}
reported an integrated flux density of 1.68$\pm$0.09~mJy with a
deconvolved diameter of less than 550~mas (12~pc) on 1999 Aug 29 UT,
and a spectral index $\alpha = -0.60\pm0.08$ ($S \propto
\nu^{\alpha}$) between 1.4 and 4.9~GHz.  The HSA imaging thus recovers
about 0.4 of the flux density available at 1.4~GHz.

\section{Implications and Future Directions}

The IMBH that energizes the Seyfert nucleus in NGC\,4395 has an
Eddington ratio of only $\sim 1.2 \times 10^{-3}$ \citep{pet05}.  By
analogy with supermassive black holes and stellar-mass black holes,
such an Eddington ratio suggests that this IMBH is in a low/hard or
quiescent state and, thus, is expected to launch a steady polar
outflow \citep{ho05,jes05,nag05,nar05}.  The elongated structure in
Figure~1 is indeed suggestive of a radio outflow on sub-parsec scales
from the IMBH in NGC\,4395.  This radio structure extends for about
15~mas (0.3~pc) along a PA of about 28\arcdeg.  From the observed
smoothness of the H$\alpha$ line, \citet{lao06} suggest that the BLR
gas is structured as a smooth, rotationally-dominated flow, most
likely in a geometrically thick configuration.  This geometrically
thick structure may be involved in launching the suspected radio
outflow.

For perspective, 20 of the 44 low-luminosity AGN (LLAGN) surveyed at
mas resolution by \citet{nag05} do show structure on sub-parsec
scales.  Six of those 20 have sub-parsec scale structures with weak or
no known larger scale jets, perhaps because the jet does not propagate
beyond the inner few parsecs.  For NGC\,4395 at 1.4~GHz, the VLA
source lacks adjacent emission and has the same integrated flux
densities for deconvolved sizes of 10-40~pc, whereas the HSA imaging
traces a suspected outflow on sub-parsec scales.  Given these VLA and
HSA properties, NGC\,4395 can now be added to the subset of LLAGN
whose emission does not extend beyond the inner few parsecs.  The
spectral power of the emission recovered in the left panel of Figure~1
is $P_{1.4} = 1.6 \times 10^{18}$~W~Hz$^{-1}$, while that for the
corrected peak in the right panel of Figure~1 is $P_{1.4}= 5.3 \times
10^{17}$~W~Hz$^{-1}$.  As expected, these powers bracket the value
derived from the VLBA discovery image \citep{wro01}.  The sub-parsec
scale of the suspected outflow in NGC\,4395 corresponds to 10$^7$
Schwarszchild radii, very different from the upper limits of 10$^3$ to
10$^4$ Schwarszchild radii inferred for several LLAGN with flat radio
spectra \citep{and04}.

The potential importance of outflows in mediating black-hole growth
has been highlighted by \citet{pel05}, motivating an estimate of the
kinetic luminosity of the suspected outflow from NGC\,4395.  But
unless the environs of the LLAGN are spatially resolved, such
estimates can be very model-dependent.  The environs of the LLAGN in
M51, at a distance of 8.4~Mpc, are spatially resolved in the radio,
optical, and X-ray regimes \citep{ter01}.  Moreover, VLA imaging at
1.4~GHz shows that the LLAGN in M51 is only eight times as powerful as
the LLAGN in NGC\,4395, and similarly compact \citep{ho01}.  For M51,
\citet{ter01} estimate that the kinetic luminosity of its jet is
comparable to the bolometric luminosity of its LLAGN.  For NGC\,4395,
the bolometric luminosity of its LLAGN is $\sim 5.4 \times
10^{40}$~ergs~s$^{-1}$ \citep{pet05}, so by analogy with M51, a
kinetic luminosity of this order could be available within the inner
few parsecs of NGC\,4395.  Approximating the radio luminosity of
NGC\,4395 as the product of the frequency times power, the luminosity
of the emission recovered in the left panel of Figure~1 is $L_{1.4} =
2.4 \times 10^{34}$~ergs~s$^{-1}$, while that for the corrected peak
in the right panel of Figure~1 is $L_{1.4} = 7.6 \times
10^{33}$~ergs~s$^{-1}$.  Although these radio luminosities are about
six orders of magnitude below the bolometric luminosity of the Seyfert
nucleus, the HSA imaging helps set a characteristic length scale for
the available kinetic luminosity.

The absorption-corrected, time-averaged 2-10 KeV luminosity of the
Seyfert nucleus in NGC\,4395 is $L_X = 8.8 \times
10^{39}$~ergs~s$^{-1}$ \citep{mor05}.  Combining this with the 5-GHz
luminosity of $L_5 = 8.8 \times 10^{34}$~ergs~s$^{-1}$ \citep{ho01},
the nucleus of NGC\,4395 is formally radio-quiet, with $log(L_5/L_X) =
-5$ \citep{ter03}.  Similarly, the nucleus is formally radio-quiet
when cast in terms of its radio-to-optical ratio \citep{gre06}.
Still, this Seyfert nucleus seems able to support a radio outflow,
suggesting that the physically defining characteristic is the
Eddington ratio of its energizing IMBH.  Hence while NGC\,4395 can be
classed as a narrow-line Seyfert 1 galaxy energized by an IMBH, its
radio properties might be expected to differ from those for a sample
of similar galaxies selected to have Eddington ratios near unity
\citep{gre06}.

The PA of the sub-parsec scale radio structure, about 28\arcdeg, can
be compared to the other nuclear structures in NGC\,4395 imaged at
lower resolution with the {\em Hubble Space Telescope} \citep[{\em
HST};][]{fil93,mat99,fil03}.  In a WFPC2 I-band image, the Seyfert
nucleus is located at the photometric center of a nuclear star cluster
with a half-light diameter of 380~mas (8.0~pc), an axis ratio of 0.83,
and a minor-axis PA of 9\arcdeg\, \citep{fil03}, confirming the
earlier findings in the I band by \citet{mat99}.  In contrast, the
galactic disk has a minor-axis PA of 54\arcdeg\, as traced by HI
\citep{swa99} and 57\arcdeg\, as traced by starlight \citep{ho01}.
The elongation PA of the sub-parsec radio structure in NGC\,4395 is
thus closer to the minor axis of the star cluster than that of the
galactic disk.  But the significance of this is unclear, given the
propensity of jet directions in Seyfert galaxies to be uncorrelated
with the minor axes of their galactic disks \citep{kin00}.

In a WFPC F502N image centered on the [OIII] $\lambda$5007 line, the
nuclear structure is somewhat resolved with emission-line gas
extending to the west of the Seyfert nucleus \citep{fil93}.  Using
their WFPC2 B-band and I-band images, \citet{mat99} do find evidence
for two structures to the west, namely a small blue arc offset by
250~mas (5.2~pc) and an extended blue plume offset by about 1\arcsec\,
(21~pc).  \citet{mat99} argue that these two blue features are
probably due to [OIII] emission from gas within a nuclear ionization
cone.  The opening angle of the cone structure covers PAs of
265-280\arcdeg, whereas the sub-parsec scale radio structure defines
PAs of about 28/208\arcdeg.  The relation between these two structures
is not presently understood, perhaps due to projection effects for
both structures.

The naturally-weighted HSA image shows hints of 5 $\sigma$ emission
about 250~mas (5.2~pc) to the west of the nucleus, thus in the
vicinity of the small blue arc reported by \citet{mat99}.  The reality
of this faint extranuclear emission at 1.4~GHz is, at present, far
from certain.  But if it is real, it has peak brightness temperatures
above 250,000~K in the vicinity of probable [OIII] emission
\citep{mat99}.  As the HSA imaging of the nucleus has recovered only
about forty percent of the flux density available at 1.4~GHz, the
faint extranuclear emission could be providing the first clues about
the location of the approximately 1~mJy still missing from the HSA
image.  HSA imaging with higher sensitivity could test the reality of
the faint extranuclear emission and should be done.  Also, {\em HST\/}
spectroscopic imaging of the small blue arc and the extended blue
plume discovered by \citet{mat99} should be done to test the
suggestion by those authors that these blue structures trace [OIII]
emission.

The nuclear star clusters in other late-type spiral galaxies are in
most respects very similar to that in NGC\,4395, and within each star
cluster a mixture of stellar populations of different ages is found,
suggesting a prolonged formation period \citep{wal06}.  For NGC\,4395,
few constraints on the cluster's stellar populations are available.
\citet{fil93} find no UV evidence for the P Cygni profiles that would
be expected from winds from massive OB stars.  \citet{fil03} do detect
the calcium infrared triplets, indicating that red supergiants are
present.  The half-light diameter of the nuclear star cluster is
380~mas (8.0~pc), so the HSA imaging samples most of the cluster's
volume.  To further constrain the stellar populations, as well as to
assist with the radio-optical registration, the naturally-weighted HSA
image was searched for cluster emitters within the half-light
diameter.  None were found above a peak flux density of 0.05~mJy per
beam, corresponding to a brightness temperature threshold of 250,000~K
and a power threshold of $P_{1.4}= 1.1 \times 10^{17}$~W~Hz$^{-1}$.
For comparison, an analog to the radio counterpart to an ultraluminous
X-ray source, at a distance of 4.8~Mpc in NGC\,5408 \citep{kar03},
would emit at the 0.3-mJy level at 4.8~GHz at the distance of
NGC\,4395.  Such emission would be consistent with the VLA photometry
of NGC\,4395 at 4.9~GHz \citep{ho01}.  However, given its
steep-spectrum nature, such a radio analog would be boosted well above
the HSA detection threshold at 1.4~GHz and, thus, cannot be present
within the star cluster's half-light diameter.  HSA imaging with
higher sensitivity would help constraint other classes of cluster
emitters.


\acknowledgments The authors thank J. Ulvestad for useful feedback on
the draft manuscript.

{\it Facilities:} \facility{GBT}, \facility{VLA}, \facility{VLBA}.

\clearpage

\begin{figure}
\epsscale{1.0}
\plottwo{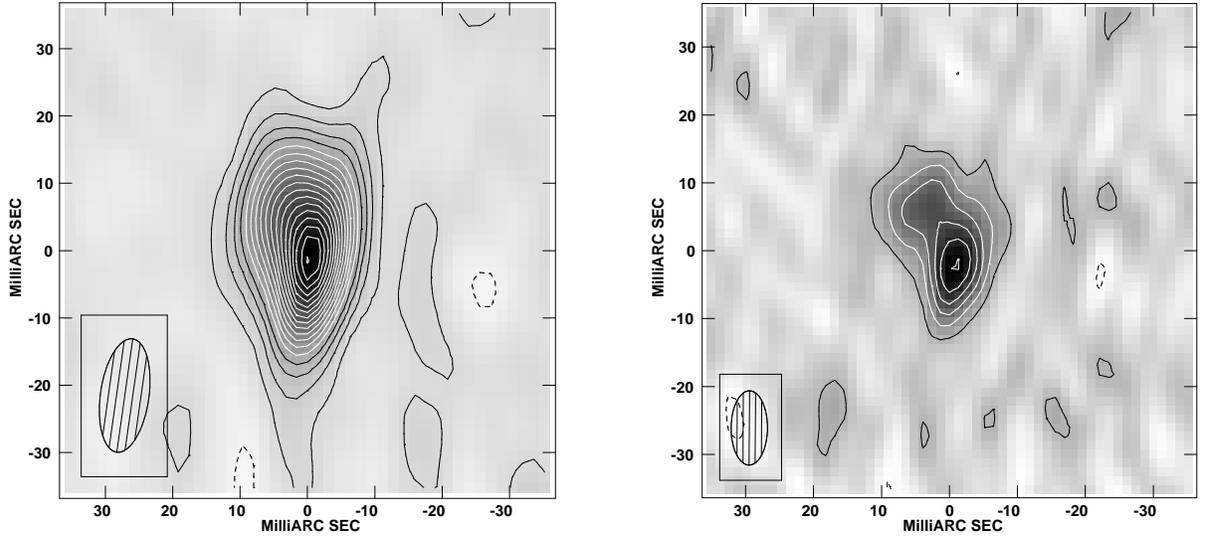}{f1b.eps}
\caption{ HSA images of Stokes $I\/$ emission from NGC\,4395 at a
frequency of 1.4~GHz, centered on the {\em a priori\/} observation
position, and spanning 70~mas (1.5~pc) per coordinate.  Negative
contours are dashed and positive ones are solid.
{\em Left:\/} Natural weighting, giving an rms noise of
0.010~mJy~beam$^{-1}$ (1 $\sigma$) and beam dimensions at FWHM of
17~mas (0.4~pc) by 7.2~mas (0.2~pc) with elongation PA -8\arcdeg\,
(hatched ellipse).  Linear grey scale spans -0.04~mJy~beam$^{-1}$ to
the peak of 0.36 ~mJy~beam$^{-1}$.  Contours are at -6, -4, -2, 2, 4,
6, 8, 10, 12, 14, ... 36 times $\sigma$.
{\em Right:\/} Robustness parameter of -0.3, giving an rms noise of
0.019~mJy~beam$^{-1}$ (1 $\sigma$) and beam dimensions at FWHM of
11~mas (0.2~pc) by 5.4~mas (0.1~pc) with elongation PA 0\arcdeg\,
(hatched ellipse).  Linear grey scale spans -0.04~mJy~beam$^{-1}$ to
the peak of 0.23 ~mJy~beam$^{-1}$.  Contours are at -6, -4, -2, 2, 4,
6, 8, 10, and 12 times $\sigma$.}
\label{fig1}
\end{figure}

\end{document}